\documentclass[twocolumn,showpacs,aps,epsfig]{revtex4}

\ifx\pdfoutput\undefined
\usepackage{graphicx}
\else
\usepackage[pdftex]{graphicx}
\usepackage{epstopdf}
\fi

\usepackage[center]{subfigure}

\begin{document}

 \newcommand{\bq}{\begin{equation}}
 \newcommand{\eq}{\end{equation}}
 \newcommand{\bqn}{\begin{eqnarray}}
 \newcommand{\eqn}{\end{eqnarray}}
 \newcommand{\nb}{\nonumber}
 \newcommand{\lb}{\label}

\title{Black hole formation from collapsing dust fluid in a 
background of dark energy }
\author{ Rong-Gen Cai ${ }^{1}$ \thanks{E-mail address: cairg@itp.ac.cn} 
and Anzhong Wang ${ }^{2}$ 
\thanks{ E-mail address: Anzhong$\_$Wang@baylor.edu}}
\affiliation{ ${ }^{1}$ Institute of Theoretical Physics, Chinese
Academy of Sciences, P.O. Box 2735, Beijing 100080, China\\
${ }^{2}$ CASPER, Physics Department, Baylor University,
 101 Bagby Avenue, Waco, TX76706}
  
\date{\today }

\begin{abstract}

The gravitational collapse of a spherically symmetric star, made of a
dust fluid, $\rho_{DM}$, in a background of dark energy, $p = w\rho,\; 
(w < -1/3)$, is studied. It is found that when only dark energy  is 
present, black holes are {\em never} formed.  When both of them 
are present, black holes can be formed, due to the condensation of the 
dust fluid. Initially the dust fluid may not play an important role, 
but, as time increases, it will dominate the collapse and finally leads 
to formation of black holes. This result remains true even when the 
interaction between the dust fluid and dark energy does not vanish. 
When  $w < -1$ (phantoms), some models can also be interpreted as 
representing the death of a white hole that ejects both dust   
and phantoms. The ejected matter re-collapses to form a black hole.

\end{abstract}

\vspace{.7cm}


\maketitle

\section{Introduction}

\renewcommand{\theequation}{1.\arabic{equation}}
\setcounter{equation}{0}

Over the past decade, one of the most remarkable discoveries is  
that our universe is currently accelerating. This was first 
observed from high red shift supernova Ia \cite{snae}, and  
confirmed later by cross checks from the cosmic microwave 
background radiation \cite{wmap} and large scale structure 
\cite{sdss}.

In Einstein's general relativity, in order to have such an 
acceleration, one needs to introduce a component to the matter
distribution of the universe with a large negative pressure. This
component is usually referred to as  dark energy. Astronomical
observations indicate that our universe is flat and currently 
consists of approximately $2/3$ dark energy and $1/3$ dark 
matter. The nature of dark energy as well as dark matter is 
unknown, and many radically different models have been proposed, 
such as, a tiny positive cosmological constant, quintessence, 
phantoms, Chaplygin gas, and dark energy in brane worlds, among 
many others [See the review articles 
\cite{SS00,Car01,PR02,Pad03,Sah04,Sah05}, and references therein].

On the other hand, another very important issue in gravitational 
physics is black holes and their formation in our universe. Although 
it is generally believed that on scales much smaller than the horizon 
size the fluctuations of dark energy itself are unimportant \cite{Ma99}, 
their effects on the evolution of matter overdensities may be 
significant \cite{FJ97}. Then, a natural question is how dark
energy affects the process of the gravitational collapse of a  
star. It is known that dark energy exerts a repulsive force
on its surrounding, and this repulsive force may 
prevent the star from collapse. Indeed, there are speculations that 
{\em a massive star doesn't simply collapse to form a black hole, 
instead,  to the formation of stars that contain dark energy. 
As a result, black holes may not exist at all} \cite{DEStar}. 
Another related issue is that how dark energy affects 
already-formed black holes (if they indeed exist in our universe).
Recently, it was shown that the mass of a black hole decreases due 
to phantom energy accretion and tends to zero when the Big Rip 
approaches \cite{BDE04}. 

In this paper, we shall study the formation of black holes from  
the gravitational collapse of a dust cloud in the background of dark 
energy, here ``dust cloud" means a cloud made of matter with zero pressure.
Thus, it includes the dark matter as a particular case.
In particular, in Sec. II we consider the collapse
of a homogeneous and isotropic star with finite radius, and
develop  the general formulas for the problem.
The formation of black holes is identified by the development of
apparent horizons. In Sec. III we consider the gravitational collapse
of dark energy and dust cloud separately, in order to study the
different roles that they may play during the collapse.  We 
show explicitly that the collapse of the dark energy alone can never 
form black holes. In Sec. IV, we study the collapse of a dust cloud
in the presence of dark energy, but there is no interaction between
them except for the gravitational one. It is found that black hole
can be formed due to the condensation of the dust cloud. 
In Sec. V, we study the collapse of a dust cloud and dark energy 
when the interaction between them does not vanish. We find that such
interaction does not change the output significantly. In particular, 
black hole can still be formed.  The paper is closed with Sec. VI, 
where our main conclusions are presented.

\section{Field Equations For Collapsing Spherical Star of a dust cloud}

\renewcommand{\theequation}{2.\arabic{equation}}
\setcounter{equation}{0}

In this section, we consider the gravitational collapse of a spherically
symmetric star with finite thickness, which is made of a dust cloud in
a background of dark energy. Let us first divide the spacetime into 
three different regions, $\Sigma$ and $V^{\pm}$, where $\Sigma$ denotes  
the surface of the star,  and $V^{-}\;$ ($V^{+}$) the interior (exterior) 
of the star [cf. Fig. 1]. For the sake of simplicity, we assume that the 
spacetime inside the star is homogeneous and isotropic,  similar to the
Oppenheimer-Synyder (OS) model \cite{OS39}, historically the first model  
for gravitational collapse. Then, the spacetime inside the star
is described by the metric 
\bq
\lb{2.1}
ds^{2}_{-} = dt^{2} - a^{2}(t)\left(dr^{2} + r^{2} d\Omega^{2}\right), 
\eq
where $d\Omega^{2} \equiv d\theta^{2} + \sin^{2}\theta d\varphi^{2}$, 
and $a(t)$ is an arbitrary function of $t$ only. Although this is a 
very ideal case, we do believe that this captures the main features of 
gravitational collapse in the background of dark energy, similar to the 
OS model that gives  most of the main properties of a collapsing star 
in an otherwise flat background \cite{Joshi}.  Since the matter fields 
are comoving in the spacetime described by metric (\ref{2.1}), 
one may choose the surface $\Sigma$ to be described by  
\bq
\lb{2.7a}
\left. r\right|_{\Sigma} =  {\mbox{Constant, say}}, \; r_{\Sigma}, 
\eq
in the $x^{-\; \mu}$-coordinates. Introducing the intrinsic coordinates 
$\xi^{a}$ on $\Sigma$ by $\xi^{a} 
\equiv (\tau, \theta, \varphi)$, the metric on $\Sigma$ can be cast 
in the form,
\bq
\lb{2.8}
\left.ds^{2}\right|_{\Sigma}  \equiv \gamma_{ab}d\xi^{a}d\xi^{b} 
= d\tau^{2} - R^{2}(\tau)d\Omega^{2},
\eq
where
\bq
\lb{2.9}
\tau = t, \;\;\; R(\tau) = r_{\Sigma}a(\tau).
\eq
Then, the normal vector $n^{-}_{\alpha}$ to the surface $\Sigma$ is 
given by,
\bq
\lb{2.9a}
n^{-}_{\alpha} = \frac{1}{a(t)}\delta^{r}_{\alpha},
\eq
and the non-zero components of the corresponding extrinsic curvature, defined by 
\bq
\lb{2.9b}
K_{ab} = - n_{\alpha}\left(\frac{\partial^{2}x^{\alpha}}
{\partial \xi^{a} \partial \xi^{b}}
+ \Gamma^{\alpha}_{\beta\delta} \frac{\partial x^{\beta}}
{\partial \xi^{a}}\frac{\partial x^{\delta}}
{\partial \xi^{b}}\right),
\eq
are given by,
\bq
\lb{2.9c}
K^{-}_{\theta\theta} = \sin^{-2}\theta K^{-}_{\phi\phi}
= r_{\Sigma} a(t).
\eq

On the other hand,  the metric outside the collapsing cloud in general can be 
cast in the form,  
\bq
\lb{2.1a}
ds^{2}_{+} = A^{2}\left(T, R\right) dT^{2} - B^{2}\left(T, R\right)\left(dR^{2}
+    R^{2} d\Omega^{2}\right), 
\eq
where  $x^{+\mu} \equiv \{T, R, \theta, 
\phi\}$ denote the coordinates used outside of the collapsing dust cloud.  The surface 
$\Sigma$ in the $x^{+\mu}$-coordinates  can be
expressed as 
\bq
\lb{2.1b}
R = R_{0}\left(T\right),   
\eq
for which the normal vector to $\Sigma$ is given by
\bq
\lb{2.1c}
n^{+}_{\alpha} = \frac{AB}{\sqrt{A^{2} - {{R'}_{0}}^{2}B^{2}}}
\left\{\delta^{R}_{\alpha} - {R'}_{0}\delta^{T}_{\alpha}\right\},
\eq
where ${R'}_{0} \equiv dR_{0}/dT$. Then, the junction conditions
$\left. ds^{2}_{-}\right|_{\Sigma^{-}} = \left. ds^{2}_{+}\right|_{\Sigma^{+}}$
require
\bqn
\lb{2.1da}
\frac{dT}{dt} &=& \frac{1}{\sqrt{A^{2}  
- {{R'}_{0}}^{2}B^{2}}}, \\
\lb{2.1db}
r_{\Sigma}a(t) &=& R_{0}(T) B\left(T, R_{0}(T)\right).
\eqn
The non-vanishing components of the extrinsic curvature, $K^{+}_{ab}$, are
given by,
\bqn
\lb{2.1ea}
K^{+}_{\tau\tau} &=&  \frac{AB}{\left(A^{2} - {{R'}_{0}}^{2}B^{2}\right)^{3/2}}
\left\{\frac{BB_{,T}}{A^{2}}{{R'}_{0}}^{3} \right.\nb\\
& &  + \left(2\frac{A_{,R}}{A} - \frac{B_{,R}}{B}\right){{R'}_{0}}^{2}
+ \left(\frac{A_{,T}}{A} - 2\frac{B_{,T}}{B}\right){{R'}_{0}} \nb\\
& & \left. - R''_{0} -  \frac{AA_{,R}}{B^{2}}\right\},\\
\lb{2.1eb}
K^{+}_{\theta\theta} &=& \sin^{-2}\theta K^{-}_{\phi\phi}
= \frac{AB^{2}R_{0}}{\left(A^{2} - {{R'}_{0}}^{2}B^{2}\right)^{1/2}}\nb\\
& & \times \left(\frac{R_{0}B_{,T}}{A^{2}}{R'}_{0}
+ \frac{B + R_{0}B_{,R}}{B^{2}}\right).
\eqn

Depending on the choice
of the spacetime outside the star, there are two possibilities \cite{Israel66}:
one is that the induced metrics $\gamma_{ab}^{\pm}$ 
are continuous across  $\Sigma$, but not $K_{ab}^{\pm}$. 
In this case, one can show that  $\Sigma$ is an energy layer,
and an infinitely thin matter shell appears on $\Sigma$.
The other possibility is that both $\gamma_{ab}^{\pm}$ 
and $K_{ab}^{\pm}$ are continuous across  $\Sigma$. 
In this case, Israel junction conditions tell us that no such a
shell appears on $\Sigma$. Once the spacetime inside the surface
is fixed, that whether a thin shell appears on $\Sigma$ or not is
completely determined by the spacetime outside the star. 
In this paper, we shall consider only the latter case, that is,
\bqn
\lb{2.1fa}
K^{+}_{\tau\tau} &=& K^{-}_{\tau\tau},\\
\lb{2.1fb}
K^{+}_{\theta\theta} &=& K^{-}_{\theta\theta}.
\eqn
Eqs.(\ref{2.1da}), (\ref{2.1db}), (\ref{2.1fa}) and (\ref{2.1fb})
are the four equations that the four functions $T(t)$, $\; R_{0}(T(t))$,
$\; A\left(T(t),R_{0}(T(t))\right)$ and $B\left(T(t),R_{0}(T(t))\right)$
must satisfy on the boundary $\Sigma$. Once the dependence of $A$ and
$B$ on $T$ and $R$ are given, these equations will uniquely determine 
the time evolution of $T$, $R_{0}$, $A$ and $B$ along the hypersurface
$\Sigma$.

However, such dependence may not be always consistent with these four equations.
For example, it is well-known that if the spacetime outside the collapsing
cloud is empty, then there is no way to do the matching for the case where
the dark energy does not vanish ($p \not= 0$) \cite{Santos85}. An interesting case is
that the spacetime outside the dust cloud is described by the McVittie solutions
\cite{McV}, which in general describes a Schwarzschild black hole in the 
background of the FRW cosmology. This is under our current investigations \cite{ZCW05}.

In this paper we shall focus our attention mainly in the 
spacetime inside the star. The main reasons for doing so are two-fold.
First, such a matching, in general, is not unique, as shown above. In principle, 
there are infinite ways to do the matching. Second, if the collapsing 
star finally forms a black hole, an apparent horizon {\em must} 
develop inside the star, and there  exists a moment 
at which the whole star collapses inside the apparent horizon.    
Clearly, this moment is exactly $\tau_{AH}$ [cf. Fig. 1], given below
by Eq.(\ref{2.12}). On the other hand, if the collapse doesn't form 
a black hole, apparent horizons will never be formed inside the
star, and the star will not be trapped at any moment. As
a result, the condition (\ref{2.11})  remains true all the 
time. Therefore, the analysis whether a black hole 
is formed reduces to the one whether apparent horizons develop
inside the star, that is, whether Eq.(\ref{2.12}) has real solutions
or not. It  does not depend on the matching,  neither on which
spacetime outside the star is chosen, although the matching
and spacetime outside the star do affect the total mass of the 
black hole and the global structure of the  spacetime. This
can be seen more clearly in the  model to be presented in Sec.
III. 1, which is essentially the OS model. As we mentioned above,
in this paper we are mainly concerned with whether or not a black 
hole can be formed from the collapse of a dust cloud in the 
background of dark energy, therefore, in the following
we shall consider only the spacetime inside the star.

 \begin{figure}[htbp]
 \begin{center}
 \label{fig0}
 \leavevmode
 \includegraphics[width=\columnwidth]{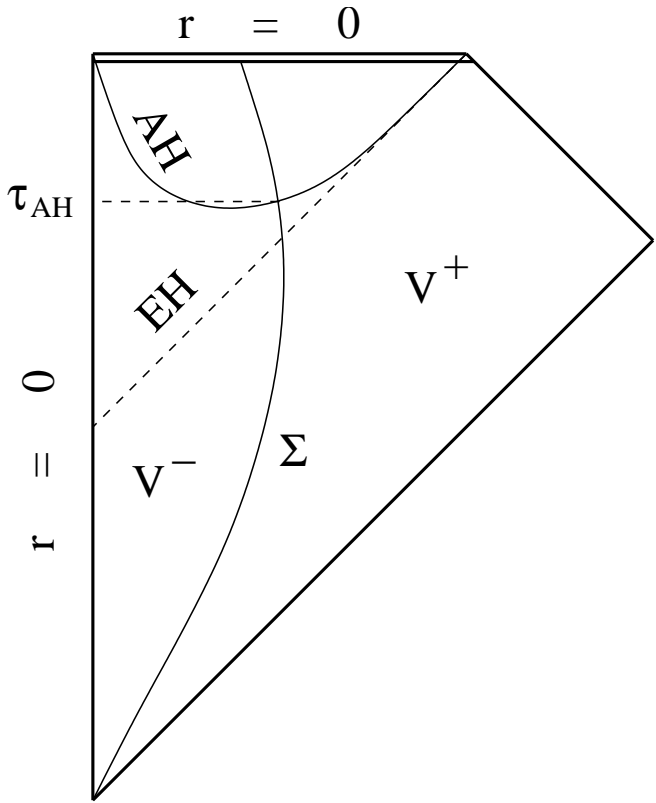} 
 \caption{The Penrose diagram of a collapsing star with a finite thickness
that finally forms a
black hole. The line $\Sigma$ represent the history of the surface of the 
star. The curved line $AH$ denotes the location of the apparent horizon,
while the dashed line $EH$ the location of the event horizon, which is 
known only at the end of the collapse. The crossing point, 
$\tau = \tau_{AH}$, denotes the moment when the whole star collapses inside 
the apparent horizon. Then, $M(\tau_{AH})$, defined by Eq.(\ref{2.10}),
represents the total contribution of the internal part
of the star to  the  total mass  of the black hole.} 
 \end{center} 
 \end{figure} 

The energy-momentum tensor (EMT) $T^{-}_{\mu\nu}$ inside the star
is given by 
\bq
\lb{2.2}
T_{\mu\nu}^{-} = \left(\rho_{DM} + \rho + p\right)u^{-}_{\mu}u^{-}_{\nu}
- p g_{\mu\nu}^{-},
\eq
where $\rho_{DM}$ denotes the energy density of the dust cloud, and 
$\rho$ and $p$ are, respectively, the energy density and pressure 
of the dark energy, while $u^{-}_{\mu}$ is their four-velocity. Since 
the fluid is comoving with the coordinates, we have
$u^{-}_{\mu} = \delta^{t}_{\mu}$. Then, the Einstein field 
equations $G_{\mu\nu}^{-} = \kappa T_{\mu\nu}^{-}$ read,
\bqn
\lb{2.3a}
& & \frac{\ddot{a}}{a}  = - 
\frac{1}{6}\kappa\left(\rho_{DM} + \rho + 3p\right),\\
\lb{2.3b}
& & \frac{{\dot{a}}^{2}}{a^{2}} = \frac{1}{3}\kappa\left(\rho_{DM}
+ \rho\right),
\eqn
where $\dot{a} \equiv da(t)/dt$.
The interaction between the dust cloud and dark energy 
is given by the conservation law, $T^{-}_{\mu\nu;\lambda}g^{-\; \nu\lambda}
= 0$, which in the present case reads
\bqn
\lb{2.4a}
& & \dot{\rho}_{DM} + 3\left(\frac{\dot{a}}{a}\right){\rho}_{DM} = Q,\\
\lb{2.4b}
& & \dot{\rho}  + 3\left(\frac{\dot{a}}{a}\right)\left(\rho + p\right) = -Q,
\eqn
where $Q = Q(t)$ denotes the interaction between the dust cloud and dark 
energy. Since in this paper we are mainly concerned with gravitational 
collapse, we assume that
\bq
\lb{2.5}
\dot{a} < 0.
\eq
The formation of a black hole is identified by the development of an apparent
horizon (AH), on which we have
\bq
\lb{2.6}
R_{,\alpha} R_{,\beta}g^{-\; \alpha\beta} = \left(r\dot{a}\right)^{2} - 1 = 0,
\eq
where $(\;)_{,x} \equiv \partial (\;)/\partial x$, and 
\bq
\lb{2.6a}
R(t, r) \equiv r a(t),
\eq
denotes the geometrical radius of the two-spheres, $t, r = Const.$

Another important quantity to describe the collapse is the mass function 
$m(t,r)$, defined by
\bq
\lb{2.7}
m(t,r) \equiv \frac{1}{2}R\left(1 + R_{,\alpha} R_{,\beta}g^{\alpha\beta}\right) 
= \frac{1}{2}r^{3}a {\dot{a}}^{2},
\eq
which can be interpreted as the total mass inside the radius $r$ at the
moment $t$. This definition was  first introduced by Cahill and McVittie  
\cite{CM73},  and has been widely used since then \cite{PI90}. In 
asymptotically flat spacetimes, it gives the correct Bondi mass at infinity 
\cite{Bondi}. On the surface $r = r_{\Sigma}$, Eq.(\ref{2.7}) gives 
the total mass of the collapsing star at the moment $\tau$ 
\cite{Note},
\bq
\lb{2.10}
M(\tau) \equiv m\left(r_{\Sigma}, \tau\right) 
= \frac{1}{2}R(\tau) \dot{R}^{2}(\tau).
\eq
Assuming that the time when the whole star collapses  inside
the apparent horizon  is $\tau_{AH}$ [cf. Fig. 1], 
from Eq.(\ref{2.6}) we have
\bq
\lb{2.12}
\left.\dot{R}^{2}(\tau_{AH})\right|_{\Sigma} = 1.
\eq
Then, the total contribution of   the collapsing 
star to the mass  of the black hole is given by 
\bq
\lb{2.10a}
M_{BH} = M(\tau_{AH}).
\eq
If no matter continuously falls into the black hole from  outside of 
the star after the moment $\tau_{AH}$, we can 
see that  Eq.(\ref{2.10a}) gives the total mass of the black hole.

Since in this paper we are mainly interested in the formation of black holes due 
to the gravitational collapse of the star, we assume that at the 
initial of the collapse, $\tau = \tau_{i}$, the star is not trapped, 
that is,
\bq
\lb{2.11}
\left. R_{,\alpha} R_{,\beta}g^{\alpha\beta} \right|_{\tau = \tau_{i}}
= \left(r_{\Sigma}\dot{a}(\tau_{i})\right)^{2} - 1 < 0.
\eq

Once we have the general formulas, in the following three sections,
$III - V$, we shall consider some specific models. Although most
of the solutions are already known in the context of cosmology,
the studies of them in the context of gravitational collapse are new,
and, as we shall see below, shall lead to important conclusions 
regarding to whether black holes can be formed in the background
of dark energy.
 

\section{Gravitational Collapse of a dust cloud or Dark Energy}

\renewcommand{\theequation}{3.\arabic{equation}}
\setcounter{equation}{0}

In this section, we consider a collapsing dust cloud and dark energy 
separately, in order to see the different roles that they may play 
during the collapse.

\subsection{Gravitational Collapse of a dust cloud}

In this case we assume that
\bq
\lb{2.5a}
\rho_{DM} \not= 0, \;\;\; \rho = 0 = p.
\eq
That is, the collapsing star consists of only a dust cloud. 
Historically, this was the first example to study gravitational 
collapse, which leads to the formation of black holes \cite{OS39}.  
In the following, we shall briefly review the main properties of 
the collapse in the framework given above, so we can see clearly 
the role that the dust cloud   plays during the collapse.   
From Eq.(\ref{2.4a}) we find that $\rho_{DM} = {\rho^{0}_{DM}}/{a^{3}}$, 
where $\rho^{0}_{DM}$ is an integration constant. Then,  
Eqs.(\ref{2.3b}) and (\ref{2.5}) yield
\bq
\lb{2.14}
a(t) = a_{0}\left(t_{0} - t\right)^{2/3},
\eq
where $a_{0} \equiv \left(3\kappa\rho^{0}_{DM}/4\right)^{1/3}$, 
and $t_{0}$ 
is another integration constant. Hence, the physically relevant 
quantities in this case are given by
\bqn
\lb{2.15}
\rho_{DM} &=& \frac{4}{3\kappa\left(t_{0} - t\right)^{2}},\nb\\
\dot{R}(\tau) &=& -\frac{2}{3}R_{0}\left(\tau_{0} 
- \tau\right)^{-1/3},\nb\\
M(\tau) &=& \frac{2}{9}{R_{0}}^{3},
\eqn
where $R_{0} \equiv r_{\Sigma}a_{0}$ and $\tau_{0} = t_{0}$. 
Assuming that the star starts to collapse at the moment
$\tau = \tau_{i}$, where the condition (\ref{2.11}) holds,
that is, the star is not initially trapped,
then from Fig. 2 and the expressions of Eq.(\ref{2.15})
we can see that the star shall collapse until the moment 
$\tau = \tau_{0}$, where a spacetime singularity is finally 
formed. This singularity is not naked, because before it is
formed, an apparent horizon is already formed at $\tau_{AH}$,
where
\bq
\lb{2.16}
\tau_{AH} = \tau_{0}- \left(\frac{2R_{0}}{3}\right)^{3} 
< \tau_{0}.
\eq
Thus, in this case the collapse actually forms black holes. 
It is interesting to note that the total mass of 
the star in the present case remains constant during the 
whole  process of the collapse. This is understandable. As
shown in \cite{OS39}, in the present case the star can be
smoothly matched to the Schwarzschild vacuum solution without
a thin shell appearing on the surface of the collapsing star.
Then, we have 
$M_{BH} = 2R^{3}_{0}/9$, which  is the mass of the Schwarzschild black hole.
 \begin{figure}[htbp]
 \begin{center}
 \label{fig1}
 \leavevmode
\includegraphics[width=\columnwidth]{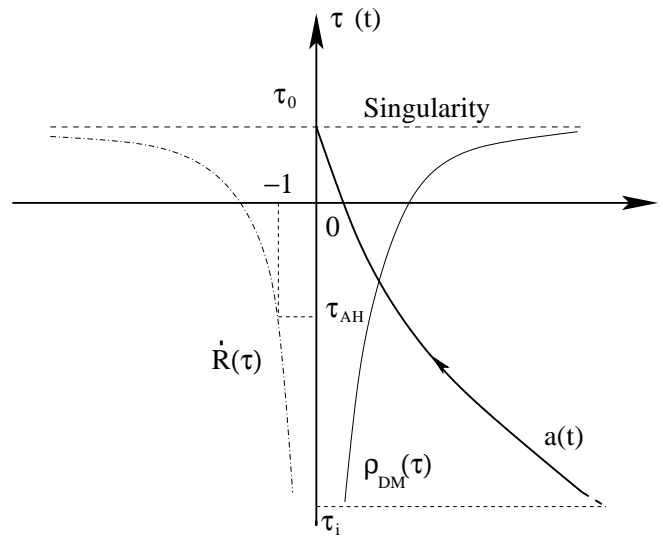}   
 \caption{The star with radius $r_{\Sigma}$, made of a dust cloud, 
starts to collapse at the moment $\tau = \tau_{i}$ in an otherwise
flat spacetime. At the moment $\tau = \tau_{AH}$ 
 an apparent horizon develops, and whereby a black hole is formed. 
 From that moment on the entire star is inside the black hole.} 
 \end{center} 
 \end{figure}

From this example, we can also see clearly that to study whether a 
collapsing star forms a black hole or not now  indeed becomes  
to study whether an apparent horizon develops in the 
internal region of the star.

\subsection{Gravitational Collapse of Dark Energy}

To study the effects of dark energy on gravitational collapse, 
 we first consider the case where
\bq
\lb{2.17aa}
\rho_{DM} = 0,\;\;\; p = w\rho \not= 0, 
\eq
where $w$ is a non-zero constant. 
When $w < -1/3$ the strong energy condition is not satisfied \cite{HE73}, 
and the fluid is said to be made of dark energy. 
It can be shown that the solution  in this case is given by
\bqn
\lb{2.17a}
a(t) &=& a_{0}\left(t_{0} - t\right)^{\frac{2}{3(1+w)}},\nb\\
\rho(t) &=& \frac{4}{3\kappa(1+w)^{2}\left(t_{0} - t\right)^{2}},\nb\\
\dot{R}(\tau) &=& - \frac{2R_{0}}{3(1+w)}\left(\tau_{0} -
\tau\right)^{-\frac{1+3w}{3(1+w)}},
\eqn
for $w > -1$, 
\bqn
\lb{2.17b}
a(t) &=& a_{0} \; {\mbox{exp}}
\left\{\left(\frac{\rho_{0}\kappa}{3}\right)^{1/2}\left(t_{0} -
t\right)\right\},\nb\\
\rho(t) &=& \rho_{0},\nb\\
\dot{R}(\tau) &=& - R_{0}\left(\frac{\rho_{0}\kappa}{3}\right)^{1/2}\nb\\
& & \times
{\mbox{exp}}\left\{\left(\frac{\rho_{0}\kappa}{3}\right)^{1/2}\left(t_{0} -
t\right)\right\},
\eqn
for $w =-1$, and 
\bqn
\lb{2.17c}
a(t) &=& a_{0} \left(t - t_{0}\right)^{\frac{2}{3(1+w)}}, \nb\\
\rho(t) &=& \frac{4}{3\kappa(1+w)^{2}\left(t - t_{0}\right)^{2}},\nb\\
\dot{R}(\tau) &=& - \frac{2R_{0}}{3(|w|-1)}\left(\tau -
\tau_{0}\right)^{-\frac{3|w|-1}{3(|w|-1)}},
\eqn
for $w < -1$.

When $w > -1/3$, for which all the energy conditions are satisfied, the
collapse is quite similar to that of the dust cloud, studied in the
last subsection. In particular, a black hole is always formed, and the 
formation of  an apparent horizon happens at,
\bq
\lb{2.18}
\tau_{AH} = \tau_{0}  
- \left(\frac{2R_{0}}{3(1+w)}\right)^{\frac{3(1+w)}{1+3w}} < \tau_{0},
\eq
where  $\tau = \tau_{0} > \tau_{AH}$ is the moment when  the spacetime 
singularity develops. The total mass of the star is given by 
\bq
\lb{2.19}
M(\tau) = \frac{2{R_{0}}^{3}}{9(1+w)^{2}}
\left(\tau_{0} - \tau\right)^{- \frac{2w}{1+w}}, 
\eq
from which we can see that $M(\tau) \rightarrow \infty$ for $w > 0$
and $M(\tau) \rightarrow 0$ for $-1/3 < w < 0$,
as the spacetime singularity at $\tau = \tau_{0}$ approaches. 

It should be noted that, although $M(\tau) \rightarrow 0$ 
for $-1/3 < w < 0$, the energy density $\rho(t) \rightarrow \infty$, as 
$\tau \rightarrow \tau_{0}$, as one can see from Eq.(\ref{2.17a}). 
Then, the spacetime is still singular at $\tau_{0}$ for $w \in(-1/3, 0)$.

When $w = -{1}/{3}$, from Eq.(\ref{2.17a}) we find $\dot{R}(\tau)
 = - R_{0}$. Thus, if the collapsing star initially is not trapped, 
it will remain so until a spacetime singularity develops at the moment 
$t =t_{0}$, which is spacelike.   It should be noted that 
in this case all the energy conditions are   satisfied, and the total mass 
of the star is given by
\bq
\lb{2.20}
M(\tau) = \frac{1}{2}{R_{0}}^{3} \left(\tau_{0} - \tau\right),
\eq
which vanishes as $\tau \rightarrow \tau_{0}$, although $\rho \simeq 
(t_{0} - t)^{-2} \rightarrow \infty$ in this limit.

When $- 1 < w < -1/3$, from Eq.(\ref{2.17a}) we find that
\bqn
\lb{2.21}
\dot{R}(\tau) &=& - \frac{2R_{0}}{3(1-|w|)} \left(\tau_{0} -
\tau\right)^{\frac{3|w| - 1}{3(1 - |w|)}},\nb\\
M(\tau) &=& \frac{2{R_{0}}^{3}}{9(1-|w|)^{2}}
\left(\tau_{0} - \tau\right)^{\frac{2|w|}{1-|w|}},\nb\\
\rho(t) &=& \frac{4}{3\kappa(1+w)^{2}\left(t_{0} - t\right)^{2}}.
\eqn
The evolution of these quantities with time $\tau$ are shown in Fig. 3.
From there we can see that if the star is not trapped 
initially, it will never become trapped in the future. If the star is trapped 
initially, it will become untrapped at the moment
\bq
\lb{2.22}
\tau_{N} = \tau_{0}  
- \left(\frac{3(1-|w|)}{2R_{0}}\right)^{\frac{3(1-|w|)}{3|w| -1}}.
\eq
The  collapse always forms a spacelike singularity with 
zero mass at the moment $\tau = \tau_{0}$.  

 \begin{figure}[htbp]
 \begin{center}
 \label{fig2}
 \leavevmode
\includegraphics[width=\columnwidth]{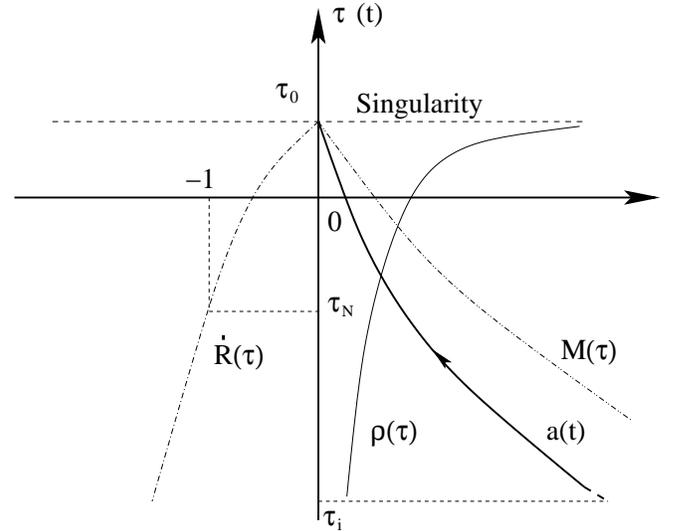} 
 \caption{The star with radius $r_{\Sigma}$, made of dark energy
with $-1 < w < -1/3$, starts to collapse at the moment 
$\tau = \tau_{i}$. If the star is trapped  initially, it 
will become untrapped at $\tau = \tau_{N}$. 
If it is not trapped initially, it will remain so in the future. The 
collapse always develops a spacelike singularity at 
$\tau = \tau_{0}$ with zero mass.} 
 \end{center} 
 \end{figure}

When $w = -1$, the solution given by Eq.(\ref{2.17b}) represents 
the de Sitter space, and the properties of this spacetime is 
well-known \cite{HE73}, so in the following we don't consider 
it any more.

 \begin{figure}[htbp]
 \begin{center}
 \label{fig3}
 \leavevmode
\includegraphics[width=\columnwidth]{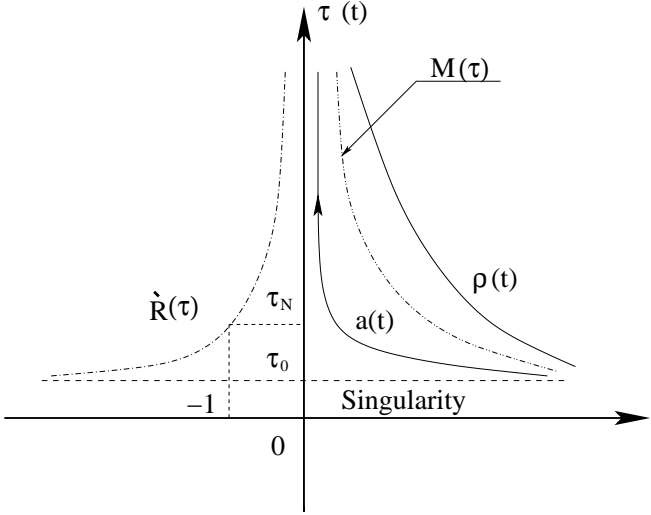}
 \caption{The star with radius $r_{\Sigma}$, made of dark energy with 
 $w < -1$, starts to collapse at the moment $\tau = \tau_{i}$.  
  The total mass of the collapsing 
star will be eventually zero, so that the spacetime is finally flat.} 
 \end{center} 
 \end{figure} 

When $w < -1$, from Eq.(\ref{2.17c}) we find that
\bq
\lb{2.23}
M(\tau) =   \frac{2{R_{0}}^{3}}{9(|w| -1)^{2}}
\left(\tau - \tau_{0}\right)^{-\frac{2|w|}{|w|-1}} 
= \cases{ 0, & $ \tau \rightarrow \infty$,\cr
\infty, & $ \tau \rightarrow  \tau_{0}$. \cr}
\eq
Some relevant quantities are plot in Fig. 4, from which  we can 
see that if $R_{,\alpha}R^{,\alpha} = \dot{R}^{2} - 1 < 0$ 
initially, it will remain so all the time. That is, in this 
case the collapse never forms a black hole, neither does a 
spacetime singularity. If it collapses initially with 
$R_{,\alpha}R^{,\alpha} = \dot{R}^{2} - 1 > 0$,  
the star will become untrapped at the moment $\tau = \tau_{N}$,
where
\bq
\lb{2.23a}
\tau_{N} = \tau_{0} 
+ \left(\frac{2R_{0}}{3(|w| - 1)}\right)^{\frac{3(|w|-1)}{3|w|-1}}.
\eq
Thus, the total mass and  energy density $\rho(t)$ of the collapsing 
star decrease as time increases, and finally become  zero in the 
limit $\tau \rightarrow \infty$. Although $a(t = \infty) = 0$,  
no spacetime singularity is formed there, as one can see from 
Eq.(\ref{2.17c}).

In review of all the above, we can see that due to its large negative 
pressure, the dark energy alone never collapses to form black 
holes.

\section{Gravitational Collapse of a dust cloud and Dark Energy:
 Without Interaction $Q = 0$}
\renewcommand{\theequation}{4.\arabic{equation}}
\setcounter{equation}{0}

When $Q = 0$, Eqs.(\ref{2.4a}) and (\ref{2.4b}) have the solutions
\bqn
\lb{2.24}
\rho_{DM} &=& \frac{\rho^{0}_{DM}}{a^{3}},\nb\\
\rho &=& \frac{\rho_{0}}{a^{3(1+w)}},
\eqn
where $\rho^{0}_{DM}$ and $\rho_{0}$ are positive constants. 
Clearly, when $w < -1$ the spacetime will be singular at both 
$a = 0$ and $a = \infty$.
From Eqs.(\ref{2.3b}) and (\ref{2.5}) we obtain
\bq
\lb{2.25}
\frac{dy}{\sqrt{1 + y^{-2w}}} = 
- \beta dt,
\eq
where
\bq
\lb{2.25a}
y \equiv \left(\frac{\rho_{0}}{\rho^{0}_{DM}}\right)^{-\frac{1}{2w}}
a^{3/2}, \;\;\;\; 
\beta \equiv \left(\frac{\rho_{0}}{\rho^{0}_{DM}}\right)^{-\frac{1}{2w}}
\left(\frac{3}{4}\kappa\rho^{0}_{DM}\right)^{1/2}.
\eq

 
\subsection{$w = -\frac{1}{2}$}
 
When $w = -1/2$, Eq.(\ref{2.25}) has the solution
\bqn
\lb{2.26}
a(t) &=& a_{0}\left(\left(t_{0} - t\right)^{2} - A^{2}\right)^{2/3},\nb\\
a_{0} &\equiv&  \left(\frac{3\kappa\rho_{0}}{16}\right)^{2/3},\;\;\;\;\;
A \equiv \left(\frac{16\rho^{0}_{DM}}{3\kappa\rho^{2}_{0}}\right)^{1/2}.
\eqn
Then, we obtain
\bqn
\lb{2.27}
\rho_{DM} &=& \frac{\rho^{0}_{DM}}{\left(\frac{3\kappa\rho_{0}}{16}\right)^{2}
\left[(t_{0} - t)^{2} - A^{2}\right]^{2}},\nb\\
\rho &=& \frac{16}{ 3\kappa \left[(t_{0} - t)^{2} - A^{2}\right]},\nb\\
\dot{R}(\tau) &=& - \frac{4}{3}R_{0}\frac{(\tau_{0} - \tau)}
{ \left[(\tau_{0} - \tau)^{2} - A^{2}\right]^{1/3}},\nb\\
M(\tau) &=& \frac{8}{9}R^{3}_{0}\left(\tau_{0} - \tau\right)^{2}.
\eqn
From the above expressions we can see that the spacetime is singular at
$t_{s}$, where $t_{s} = t_{0} - A$. We also have  
\bqn
\lb{2.28}
\dot{R}(\tau) &=& - \frac{4}{3}R_{0}\frac{(\tau_{0} - \tau)}
{ \left[(\tau_{0} - \tau)^{2} - A^{2}\right]^{1/3}}\nb\\
&=& \cases{- \infty, & $\tau \rightarrow -\infty, \; \tau_{s}$,\cr
- B, & $ \tau = \tau_{min.}$,\cr}
\eqn
where 
\bq
\lb{2.29}
\tau_{min.} \equiv \tau_{0} - \sqrt{3}A, \;\;\;\;
B \equiv  \frac{4^{2/3}}{3^{1/3}}R_{0}A^{1/3},
\eq
as shown by Fig. 5. Thus, if $B > 1$ we have $R_{,\alpha}R^{,\alpha} > 0$
all the time, and the star is trapped during the whole process of collapse.
In order to have $R_{,\alpha}R^{,\alpha} < 0$ initially, we must 
choose $r_{\Sigma},\; \rho^{0}_{DM}$ and $\rho_{0}$ such that 
\bq
\lb{2.30}
B < 1.
\eq
Once this condition   is satisfied, from Fig. 5
we can see that as long as $\tau_{i} > \tau^{-}_{AH}$, the 
collapsing star will not be trapped at the initial. However, as the time 
increases, the dust cloud becomes dominant over the 
dark energy, and an apparent horizon will finally develop at the moment
$\tau = \tau^{+}_{AH}$, where $\tau^{\pm}_{AH}$ are the two real roots
of the equation,
\bq
\lb{2.31}
\left(\tau_{0} - \tau\right)^{3} 
- \left(\frac{3}{4R_{0}}\right)^{3}\left(\tau_{0} - \tau\right)^{2}
+ A^{2}\left(\frac{3}{4R_{0}}\right)^{3}= 0,
\eq
with $\tau^{+}_{AH} > \tau^{-}_{AH}$. This can be seen clearly 
from,
\bq
\lb{2.32}
\frac{\rho_{DM}}{\rho} = \frac{16\rho^{0}_{DM}}{3\kappa\rho^{2}_{0}
\left[(t_{0} - t)^{2} - A^{2}\right]} = \cases{0, & $ t \rightarrow 
-\infty$,\cr
\infty, & $t \rightarrow t_{s}$.\cr}
\eq
Thus, a spacetime singularity develops at $t_{s}$. From Eq.(\ref{2.27}) 
we can see that the mass of a such formed black hole is finite.

\begin{figure}[htbp]
\begin{center}
\label{fig4}
\leavevmode
\includegraphics[width=\columnwidth]{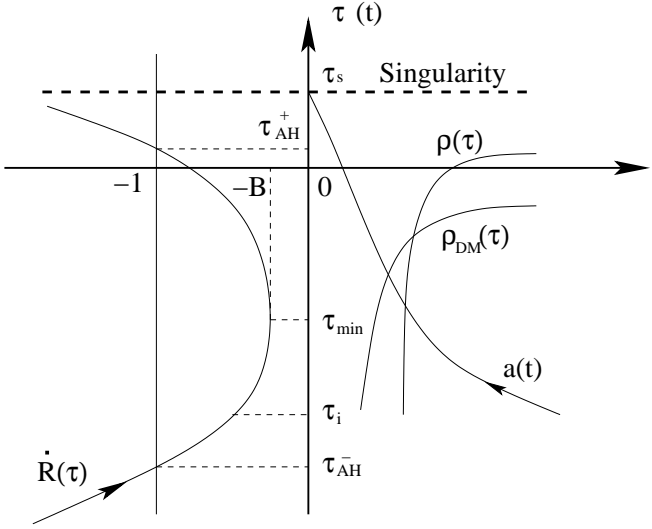}
\caption{The collapsing star with radius $r_{\Sigma}$, 
made of  a dust cloud in the background of dark energy
without interaction, $Q = 0$, for $B < 1$. It starts to collapse
at the moment $\tau = \tau_{i}$. As time increases, 
the dust cloud becomes dominant, and an apparent horizon 
finally develops at $\tau^{+}_{AH}$, whereby a black hole is 
formed. From this moment on, the star collapses entirely 
inside the black hole, and at the moment $\tau_{s}$ 
a spacetime singularity develops.   } 
 \end{center} 
 \end{figure} 

\subsection{$w = -1$}

In this case, it can be shown that the solutions are given by
\bqn
\lb{2.33}
a(t) &=& a_{0}\sinh^{2/3}
\left[\beta\left(t_{0} - t\right)\right],\nb\\
\rho_{DM} &=& \frac{\rho_{0}}{\sinh^{2}
\left[\beta\left(t_{0} - t\right)\right]},\nb\\
\rho &=& \rho_{0},
\eqn
where $a_{0}$ is a positive constant, and 
\bq
\lb{2.34}
\beta \equiv \left(\frac{3}{4}\kappa \rho_{0}\right)^{1/2}.
\eq
Then, we obtain
\bqn
\lb{2.35}
\dot{R}(\tau) &=& - \frac{2}{3} \beta R_{0}\frac{
\cosh\left[\beta\left(\tau_{0} - \tau\right)\right]}
{\sinh^{1/3}\left[\beta\left(\tau_{0} - \tau\right)\right]}\nb\\
& = &
\cases{ -\infty, & $\tau \rightarrow -\infty$,\cr
- \frac{4^{1/3}}{3^{1/2}}\beta{R_{0}}, &$ \tau = \tau_{min.}$,\cr
-\infty, & $\tau = \tau_{0}$,\cr}
\eqn
where
\bq
\lb{2.36}
\tau_{min.} \equiv  \tau_{0} - \frac{1}{\beta}\sinh^{-1}
\left(\frac{1}{\sqrt{2}}\right).
\eq
The curve of $\dot{R}(\tau)$ versus $\tau$ is quite similar to that 
given in Fig. 5, except that now the spacetime singularity occurs 
at $\tau = \tau_{0}$. Thus, in order to have the collapsing star 
untrapped at the initial, we must choose the free parameters 
$a_{0},\; r_{\Sigma}$ and $\rho_{0}$ such that
\bq
\lb{2.37}
\beta{R_{0}} < \frac{3^{1/2}}{4^{1/3}}.
\eq
Then, as shown by Fig. 5, choosing $\tau_{i} > \tau_{AH}^{-}$ we can
see that the solution can be interpreted as representing gravitational 
collapse of a dust cloud in the background of dark energy (in the 
present case it is the cosmological constant.). At initial the collapsing 
star is untrapped. However, as time increases, the dust cloud becomes 
dominant, and finally an apparent horizon  develops at the moment 
$\tau = \tau^{+}_{AH}$, whereby a black hole is formed, where 
$\tau^{\pm}_{AH}$ now are given by
\bq
\lb{2.38}
\tau^{\pm}_{AH} \equiv \tau_{0} 
- \frac{1}{\beta}\sinh^{-1}\left({X^{\pm}}^{3/2}\right),
\eq
and $X^{\pm}$ are the two real roots of the equation,
\bq
\lb{2.39}
X^{3} - \left(\frac{3}{2\beta{R_{0}}}\right)^{2}X + 1 = 0.
\eq
Note that the black hole formed in this case also has a finite non-zero mass,
 as we can see from the following expression, 
\bq
\lb{2.40}
M(\tau) = \frac{2}{9}  \beta {R_{0}}^{3}
\cosh^{2}\left[\beta\left(\tau_{0} - \tau\right)\right].
\eq

\subsection{$w < -1$}

In general, the integration of Eq.(\ref{2.25}) gives
\bq
\lb{2.41}
y \;  F\left(\frac{1}{2}, - \frac{1}{2w}; 1- \frac{1}{2w}; -
y^{-2w}\right) = -\beta\left(t - t_{0}\right),
\eq
where $F(a,b;c; z)$ denotes the ordinary hypergeometric function with
$F(a,b;c; 0) = 1$. Thus, we find   
\bq
\lb{2.42} 
y \simeq - \beta\left(t - t_{0}\right) \sim 0,
\eq
as $t \rightarrow t_{0}$. On the other hand, using the relation 
\cite{AS72},
\bqn
\lb{2.43}
F\left(a,b;c;z\right) &=& \frac{\Gamma(c)\Gamma(b-a)}
{\Gamma(b)\Gamma(c-a)}(-z)^{-a}\nb\\
& & \times F\left(a,1-c+a;1-b+a;\frac{1}{z}\right)\nb\\
& & + \frac{\Gamma(c)\Gamma(a-b)}{\Gamma(a)\Gamma(c-b)}
(-z)^{-b}\nb\\
& & \times F\left(b,1-c+b;1-a+b;\frac{1}{z}\right),\nb\\
\eqn
we find   
\bqn
\lb{2.44}
y &&  F\left(\frac{1}{2}, - \frac{1}{2w}; 1- \frac{1}{2w}; -
y^{-2w}\right) \nb\\
& & \rightarrow \pi^{-1/2}\Gamma\left(1 - \frac{1}{2w}\right)
\Gamma\left(\frac{1+w}{2w}\right),
\eqn
as $y \rightarrow \infty$ for $w < -1$. Hence, we have
\bq
\lb{2.45}
y \rightarrow \infty, 
\eq
as $t \rightarrow t_{s}$, where
\bq
\lb{2.46}
t_{s} \equiv t_{0} - \frac{1}{\beta\pi^{1/2}}
\Gamma\left(1 - \frac{1}{2w}\right)
\Gamma\left(\frac{1+w}{2w}\right).
\eq
Then, it can be seen that the curve of $a(t)$ versus 
$t$ is that given by Fig. 6.
On the other hand, from Eq.(\ref{2.25}) we also have
\bqn
\lb{2.47}
\dot{R}(\tau) &=& - R_{0}\frac{\left[\rho^{0}_{DM} 
+ \rho_{0} a^{-3w}(\tau)\right]^{1/2}}
{a^{1/2}(\tau)},\nb\\
\ddot{R}(\tau) &=&- \left(\frac{1}{12}
\kappa\right)^{1/2}\frac{R_{0}}{a^{2}}\nb\\
& & \times
\left\{\rho^{0}_{DM} - \left(3|w| - 1\right)
\rho_{0}a^{-3w}(\tau)\right\},\nb\\
M(\tau) &=& \frac{1}{2}r_{\Sigma}R^{2}_{0}
\left[\rho^{0}_{DM} + \rho_{0} a^{-3w}(\tau)\right],
\eqn
where $R_{0} \equiv (\kappa/3)^{1/2}r_{\Sigma}$. Thus, we find
\bq
\lb{2.48}
\dot{R}(\tau) = \cases{- \infty, & $\tau = \tau_{0}$,\cr
- B, & $\tau = \tau_{min.}$,\cr
- \infty, & $\tau = \tau_{s}$,\cr}
\eq
where
\bq
\lb{2.49}
B \equiv R_{0}\frac{\left(\rho^{0}_{DM} 
+ \frac{\rho^{0}_{DM}}{3|w|-1}\right)^{1/2}}
{\left(\frac{\rho^{0}_{DM}}{(3|w|-1)
\rho_{0}}\right)^{\frac{1}{6|w|}}}.
\eq
Clearly, for choice where $B < 1$, there exists initial 
moment $\tau_{i}$ for which the collapsing
star is not trapped at $\tau_{i}$. In fact, as long as 
$ \tau^{+}_{AH} > \tau_{i} > \tau^{-}_{AH}$,
the star is not trapped initially, as shown by Fig. 6, 
where $\tau^{\pm}_{AH}$ are the two 
real roots of the equation $\dot{R}^{2} - 1= 0$. 
But, the collapse will eventually develop 
an apparent horizon at $\tau^{+}_{AH}$, 
whereby a black hole is formed. The spacetime becomes
singularity at $\tau = \tau_{0}$ where $a(\tau_{0}) = 0$, 
as shown by Eq.(\ref{2.42}).  From Eq.(\ref{2.47}) we can 
see that such formed black holes have finite non-zero mass.

It is interesting to note that the solutions can also be 
interpreted as representing a white hole converting itself 
into a black hole \cite{Blau}, if we choose $\tau_{i} = 
\tau_{s}$, that is, the white hole evaporates 
through ejecting material, which will later re-collapse
to form a black hole.

 \begin{figure}[htbp]
 \begin{center}
 \label{fig5}
 \leavevmode
\includegraphics[width=\columnwidth]{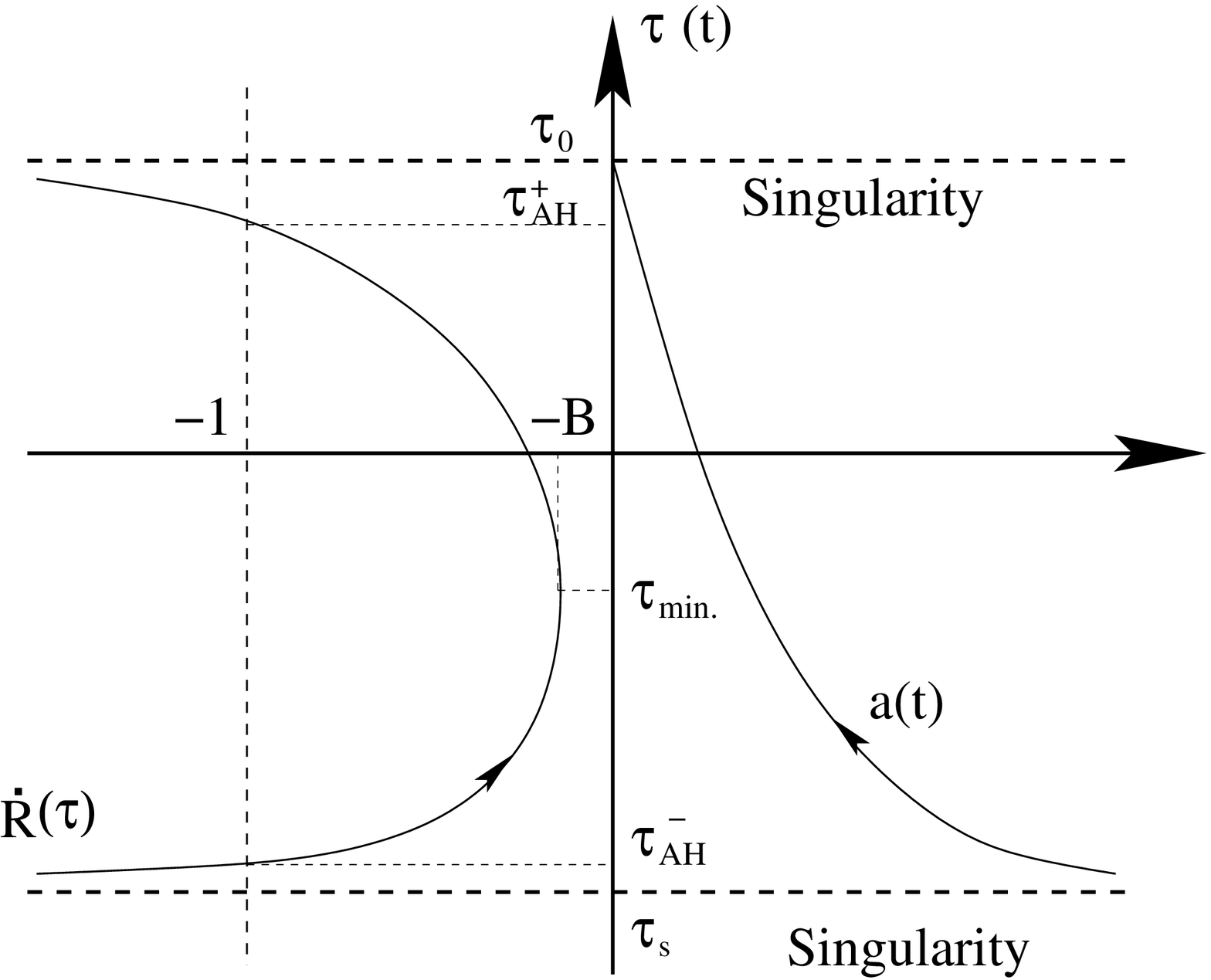}
 \caption{ The Curves of $\dot{R}(\tau)$ and $a(t)$ for 
$Q = 0$ and $w < -1$.  }
 \end{center} 
 \end{figure} 

\section{Gravitational Collapse of a dust cloud and Dark Energy: 
With Interaction $Q \not= 0$}
\renewcommand{\theequation}{5.\arabic{equation}}
\setcounter{equation}{0}

Recently, we studied the interaction of a dust cloud and dark energy in the 
context of cosmology by assuming that \cite{CW05}
\bq
\lb{5.1}
\frac{\rho}{\rho_{DM}} = A a^{3n},
\eq
where $A$ and $n$ are  two arbitrary constants, subject 
to  $A > 0$. Assuming that the dark energy satisfies the 
equation of state $p = w\rho$ with $w$ 
being a constant, from Eqs.(\ref{2.4a}), (\ref{2.4b}) and (\ref{5.1}) 
we obtain
\bqn
\lb{5.3}
\rho &=& \frac{A\rho^{0}_{t}a^{3n}}{a^{3}
\left(1 + Aa^{3n}\right)^{(w+n)/n}},\nb\\
\rho_{DM} &=& \frac{\rho^{0}_{t}}{a^{3}
\left(1 + Aa^{3n}\right)^{(w+n)/n}},
\eqn
where $\rho^{0}_{t}$ is another positive constant. 
Substituting these into Eq.(\ref{2.3b}) and considering Eq.(\ref{2.5}), 
we find 
\bq
\lb{5.4}
\left(1 + y^{2n}\right)^{\frac{w}{2n}} dy = - \beta dt,
\eq
where
\bq
\lb{5.5}
y \equiv A^{\frac{1}{2n}} a^{3/2}, \;\;\; \beta \equiv
A^{\frac{1}{2n}}\left(\frac{3}{4}\kappa\rho^{0}_{t}\right)^{1/2}.
\eq
Hence, from Eq.(\ref{2.4a}) we have 
\bq
\lb{5.6}
Q = -3A(w+n)\left(\frac{\dot{a}}{a}\right)
\frac{a^{3n}\rho_{t}}{\left(1 + Aa^{3n}\right)^{2}}.
\eq
From Eq.(\ref{5.3}), on the other hand, we obtain  
\bqn
\lb{5.6a}
\rho &= & \cases{ a^{-3(1+w)}, & $ a\rightarrow \infty$,\cr
a^{3(n-1)}, & $ a\rightarrow 0$,\cr}\nb\\
\rho_{DM} &= & \cases{ a^{-3(1+w +n)}, & $ a\rightarrow \infty$,\cr
a^{-3}, & $ a\rightarrow 0$.\cr}
\eqn
Therefore, the spacetime is always singular at $a = 0$. 
When $w < -1$, it is also singular as $a \rightarrow \infty$.

\subsection{$ n = 1/2$}

When $n = 1/2$, Eqs.(\ref{5.4}) and (\ref{5.5}) yield
\bq
\lb{5.7a}
a(t) = a_{0}\left\{\left[\beta\left(1+w\right)
\left(t_{0} - t\right)\right]^{\frac{1}{1+w}}
- 1\right\}^{2/3},
\;\;\; (w \not= -1),
\eq
for $w \not= -1$, and
\bq
\lb{5.7b}
a(t) = a_{0} \left(e^{\beta\left(t_{0} - t\right)} 
- 1\right)^{2/3},\;\;\; (w = -1),
\eq
for $w = -1$, where $a_{0} \equiv A^{-2/3}$.

When $w > -1$, we find that 
\bq
\lb{5.8}
a(t) = \cases{ \infty, & $ t \rightarrow - \infty$,\cr
0, & $t = t_{s}$,\cr}
\eq
with 
\bq
\lb{5.9}
t_{s} \equiv t_{0} - \frac{1}{\beta(1+w)}.
\eq
Then, from Eq.(\ref{5.3}) we can see that 
the spacetime is singular at
$t =t_{s}$. The nature of the singularity 
can be seen from  $\dot{R}(\tau)$,
given by
\bqn
\lb{5.10}
\dot{R}(\tau) &=& - \frac{2}{3}\beta{R_{0}} 
\frac{\left[\beta\left(1+w\right)\left(\tau_{0} 
- \tau\right)\right]^{-\frac{w}{1+w}}}
{\left\{\left[\beta\left(1+w\right)\left(\tau_{0} 
- \tau\right)\right]^{\frac{1}{1+w}}
- 1\right\}^{1/3}} \nb\\
&=&  \cases{0, & $w > -1/3$,\cr
-2\beta{R_{0}}/3, & $w = -1/3$,\cr
-\infty, & $w < -1/3$,\cr}
\eqn
as $\tau \rightarrow -\infty$. On the other hand, 
as $\tau \rightarrow \tau_{s}$
we have $\dot{R} \rightarrow - \infty$ for any 
value of $w$ with $w > -1$. 
Thus, the curve of $\dot{R}$ versus $\tau$ is 
that of Fig. 7.
 \begin{figure}[htbp]
 \begin{center}
 \label{fig6}
 \leavevmode
\includegraphics[width=\columnwidth]{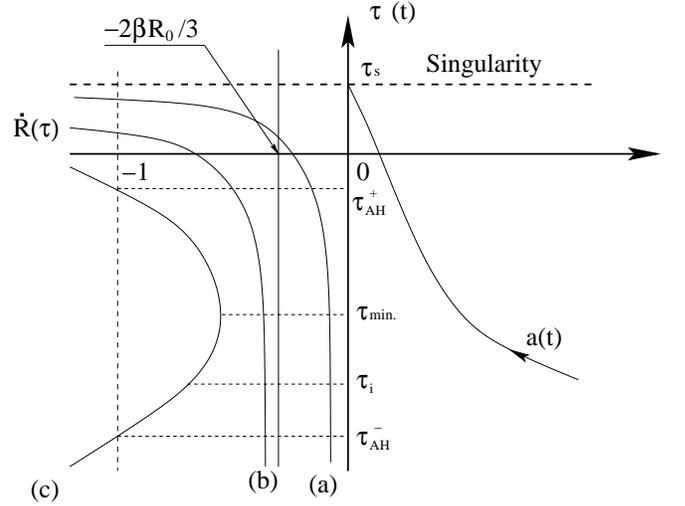}
 \caption{ The Curve of $\dot{R}(\tau)$ versus $\tau$ for 
$Q \not= 0$. (a) $\; w > -1/3$;
 (b) $\; w = -1/3$; and (c) $\; -1 < w < -1/3$.}
 \end{center} 
 \end{figure} 

When $w > -1/3$, as shown by Fig. 7, we can always choose an initial moment
where $\dot{R}(\tau_{i}) > -1$ so that the collapsing star is  not
trapped initially. As the star collapses, an apparent horizon develops at   
$\tau = \tau^{+}_{AH}$, whereby a black hole is formed. 
From this moment on, the collapsing star falls entirely inside the 
black hole. The total mass of it is finite,  
\bq
\lb{5.11a}
M(\tau) = \frac{2}{9}\beta^{2}R^{3}_{0}\left[\beta(1+w)
\left(\tau_{0} - \tau\right)\right]^{-\frac{2w}{1+w}}.
\eq

When $w = -1/3$, the main properties of the solution 
are similar to these
with $w > -1/3$, provided that $2\beta{R_{0}}/3 < 1$. When, 
$2\beta{R_{0}}/3 > 1$ 
the star will be trapped all time.

When $- 1< w < -1/3$ we find that
\bq
\lb{5.11}
\dot{R}(\tau_{min.}) = -\frac{2}{3}\beta{R_{0}}
\frac{\left(3|w|\right)^{|w|}}{\left(3|w| - 1\right)^{(3|w|-1)/3}},
\eq
with 
\bq
\lb{5.12}
\tau_{min.} = \tau_{0} - \frac{1}{\beta(1-|w|)}\left(\frac{3|w|}
{3|w| - 1}\right)^{1-|w|}.
\eq
Thus, to have the collapsing star not trapped initially, 
we must require $\dot{R}(\tau_{min.}) > -1$. Then, for any 
given moment $\tau_{i}$, where $\tau^{+}_{AH} > \tau_{i} > \tau^{-}_{AH}$, 
the collapsing star is not trapped
initially, but, as the star is  collapsing, an apparent horizon 
will develop at $\tau^{+}_{AH}$, where $\tau^{\pm}_{AH}$ are the two 
real roots of the equation $\dot{R}(\tau) = -1$.

When $w = -1$, the solution is given by Eq.(\ref{5.7b}), 
from which we find that
\bq
\lb{5.13}
a(t) = \cases{ \infty, & $ t \rightarrow - \infty$,\cr
0, & $t = t_{0}$.\cr}
\eq
From Eq.(\ref{5.3}) we obtain
\bqn
\lb{5.14} 
\rho &=& A\rho^{0}_{t} \frac{1+ Aa^{3/2}}{a^{3/2}},\nb\\
\rho_{DM} &=& \rho^{0}_{t} \frac{1+ Aa^{3/2}}{a^{3}}.
\eqn
Thus, the spacetime is  singular  at $t_{0}$, where $a(t_{0}) = 0$. On
the other hand, we also have
\bqn
\lb{5.15}
\dot{R}(\tau) &=& - \frac{2}{3}\beta{R_{0}}
\frac{e^{\beta\left(\tau_{0} - \tau\right)}}
{\left[e^{\beta\left(\tau_{0} - \tau\right)} - 1\right]^{1/3}},\nb\\
\ddot{R}(\tau) &=& - \frac{2}{9}\beta^{2}{R_{0}}
\frac{e^{\beta\left(\tau_{0} - \tau\right)}}
{\left[e^{\beta\left(\tau_{0} - \tau\right)} - 1\right]^{4/3}}
\left(3 - 2e^{\beta\left(\tau_{0} - \tau\right)}\right),\nb\\
M(\tau) &=& \frac{2}{9}\beta^{2}{R_{0}}^{3}
e^{2\beta\left(\tau_{0} - \tau\right)}.
\eqn
Then, we can see that this case is similar 
to the one for $-1 < w < -1/3$. In
particular, $\dot{R}(\tau)$ has a maximal  
 at $\tau_{min.}$, where
\bqn
\lb{5.16}
\dot{R}(\tau_{min.}) &=& -  2^{1/3}\beta{R_{0}},\nb\\
\tau_{min.} &\equiv& \tau_{0} - \frac{1}{\beta}\ln\left(\frac{3}{2}\right).
\eqn
Thus, by properly choosing the free parameters, 
the solution can be interpreted
as representing the gravitational collapse of 
a dust cloud in the presence of dark 
energy, in which the collapse will finally lead 
to the formation of black holes.

When $w < -1$, the solution is that of Eq.(\ref{5.7a}), 
which can be written
as
\bqn
\lb{5.17}
a(t) &=& a_{0}\frac{\left\{1 - \left[\beta\left(|w| - 1\right)\left(t -
t_{0}\right)\right]^{\frac{1}{|w| -1}}\right\}^{2/3}}
{\left[\beta\left(|w| - 1\right)\left(t -
t_{0}\right)\right]^{\frac{2}{3(|w| -1)}}}\nb\\
&=& \cases{0, & $ t = t_{s}$,\cr
\infty, & $t = t_{0}$,\cr},
\eqn
where
\bq
\lb{5.18}
t_{s} = t_{0} + \frac{1}{\beta(|w| - 1)}.
\eq
From Eq.(\ref{5.3}) we find that the spacetime is 
singular at both $t_{0}$
and $t_{s}$, while from Eq.(\ref{5.17}) we obtain
\bqn
\lb{5.19}
\dot{R}(\tau) &=& - \frac{2}{3}\beta{R_{0}}
\frac{\left[\beta\left(|w| - 1\right)\left(\tau -
\tau_{0}\right)\right]^{\frac{1-3|w|}{3(|w| -1)}}}{
\left\{1 - \left[\beta\left(|w| - 1\right)\left(\tau -
\tau_{0}\right)\right]^{\frac{1}{|w| -1}}\right\}^{1/3}}\nb\\
&=& \cases{ -\infty, & $ \tau = \tau_{s}$,\cr
- B, & $ \tau = \tau_{min.}$,\cr
-\infty, & $ \tau = \tau_{0}$,\cr}
\eqn
where
\bqn
\lb{5.20}
B &\equiv& - \frac{2}{3}\beta{R_{0}} 
\frac{\left(3|w|\right)^{|w|}}{\left(3|w| -
1\right)^{|w| -1/3}},\nb\\
\tau_{min.} &=& \tau_{0} + \frac{1}{\beta\left(|w| -1\right)}
\left(\frac{3|w|-1}{3|w|}\right)^{|w| -1}.
\eqn
The curve of $\dot{R}$ is that given in Fig. 6, but now with 
$\tau_{0}$ and $\tau_{s}$ being exchanged, as
in the present case we have $\tau_{0} < \tau_{s}$. If $B < 1$, 
the solutions can be interpreted as representing
the gravitational collapse of a dust cloud in the background of
phantoms, starting from a moment
$\tau_{i}$, where $\tau_{i} > \tau^{-}_{AH}$. 
The collapse develops 
an apparent horizon at $ \tau^{+}_{AH}$, 
whereby a black hole is formed.
The total mass of the collapsing star now is given by
\bq
\lb{5.20a}
M(\tau) =   \frac{2}{9}\beta^{2}R^{3}_{0}
\left[\beta\left(|w| - 1\right)
\left(\tau - \tau_{0}\right)\right]^{-\frac{2|w|}{|w| -1}},
\eq
which is finite and non-zero at $\tau_{s}$, 
when a spacetime singularity is formed.

Similar to the case where $Q = 0$ and $w < -1$, the 
solutions can also be interpreted as representing a white hole 
converting itself into a black hole \cite{Blau}.


\subsection{$n = 1$}

In this case, Eq.(\ref{5.4}) reads
\bq
\lb{5.21}
\frac{dy}{\left(1+y^{2}\right)^{m}} = - \beta dt,
\eq
where $m \equiv - w/2$. When $m = 1/2$ or $w = -1$, from Eq.(\ref{5.6}) we
find that $Q = 0$. Thus, this is the case studied in the last section. When
$m = 1$,   Eq.(\ref{5.21}) has the solution,
\bq
\lb{5.22}
a(t) = A^{-1/3} y^{2/3} = 
a_{0} \tan^{2/3}\left[\beta\left(t_{0} - t\right)\right].
\eq
It can be shown that this case is quite 
similar to the previous case $n = 1/2$ 
and $w < -1$. In particular, the curve of 
$\dot{R}(\tau)$ is quite similar to that
given by Fig. 6. Therefore, the 
solution in this case can also be interpreted 
as representing the gravitational
collapse of a dust cloud in the background of phantoms, 
in which a black hole is finally
formed. 

\section{Conclusions}

\renewcommand{\theequation}{6.\arabic{equation}}
\setcounter{equation}{0}

In this paper we studied the gravitational collapse of a spherically symmetric
star with finite radius, which is made of homogeneous and isotropic fluid. 
When the fluid has only one component with the equation of state $p = w \rho$,
we showed explicitly in Sec. III that the collapse {\em always} forms black 
holes for $w > -1/3$, including the case of a dust cloud where $w = 0$. When 
$w \le -1/3$ the collapse  {\em never} forms black holes.

In Sec. IV, we considered the collapse of the fluid that consists of two
different components, the dust cloud, $\rho_{DM}$, and the dark energy
$p = w \rho$, but assuming that, except for their gravitational interaction, 
 there is no other interaction between them. We found that black holes can 
still be formed, due to the condensation of the dust cloud. At the beginning 
of the collapse, the dust cloud may not play an important role. But, as the 
time increases, it will dominate the collapse, so that a black hole is 
finally formed.  

To study the effects of the interaction between dust cloud and dark energy,
in Sec. V we studied the gravitational collapse by assuming that  \cite{CW05}
\bq
\lb{6.1}
\frac{\rho}{\rho_{DM}} = A a^{3n},
\eq
where $A$ and $n$ are  arbitrary constants. In this case, the interaction is
characterized by [cf. Eq.(\ref{5.6})]
\bq 
\lb{6.2}
Q = -3A(w+n)\left(\frac{\dot{a}}{a}\right)
\frac{a^{3n}\rho_{t}}{\left(1 + Aa^{3n}\right)^{2}}.
\eq
By considering several specific models, we found similar conclusions as
in the case where the interaction vanishes, that is, black holes can still be 
formed due to the collapse of the dust cloud in the background of dark 
energy. 

When $w < -1$ (phantoms) some models may also be interpreted as the death 
of a white hole \cite{Blau}, that is, a white hole evaporates through 
ejecting material, which will later re-collapse to form a black hole.

Our results obtained in this paper do not seemingly support the 
speculations that black holes do not exist due to the presence   
of dark energy. Instead, due to the local condensation of the dust cloud,
black holes can still be formed even in the background of dark energy.
We believe that this is true not only for a dust cloud but also for
other matter fields that satisfy the energy conditions \cite{HE73}. 
 
Certainly, one may argue that a collapsing star that 
consists of homogeneous and isotropic fluid is a very ideal case, and 
in more realistic cases the internal region of the collapsing
star should be inhomogeneous. However, as argued previously, 
we believe that the main properties of the present model should 
remain valid even in more realistic cases. 

In addition, in this paper we did not consider the junction of 
the star to the spacetime outside of it. The main reason is that 
if apparent horizon develops inside the star, 
the collapse must form a black hole. Of course, different junctions 
will result in different masses of black holes and different 
global structures of the spacetime. The investigations of these
problems will be reported in another occasion \cite{ZCW05}.

\section*{Acknowledgments}

RGC  was supported by a grant from Chinese Academy of Sciences,
grants from NSFC, China (No. 10325525 and No.90403029), and a grant from
the Ministry of Science and Technology of China (No. TG1999075401).



\begin{thebibliography}{100}
\bibitem{Note} This gives the total mass of only the internal part of the 
star and does not include the part of the thin shell,
which usually appears on the surface of the star. In the rest of this 
paper, the total mass of the star will always refer to the
total mass of the internal part of the star.
\end{thebibliography}
\end{document}